\documentclass[cameraready]{Interspeech}

\usepackage[T1]{fontenc}
\usepackage{tipa}
\usepackage{booktabs}
\usepackage{graphicx}
\usepackage{csquotes} % gives \enquote{...}
\usepackage{tabularx}
\usepackage{array}
\usepackage{makecell}

\newcommand{\ipatok}[1]{\textipa{[#1]}}  % bracketed token
\newcommand{\ipa}[1]{\textipa{#1}}       % inline IPA

\title{Phoneme-Level Mispronunciation Screening in Polish-Speaking Children with an Explainable Assistant}

\author[affiliation={1, 2}, orcid=0009-0008-4299-9162, correspondingauthor]{Milosz}{Dudek}
\author[affiliation={1, 2}, orcid=0000-0002-2193-7690]{Daria}{Hemmerling}
\author[affiliation={1, 2}, orcid=0000-0002-1392-4291]{Kamil}{Kwarciak}
\author[affiliation={2}]{Maciej}{Stroinski}
\author[affiliation={2}]{Maria}{Pensko}
\author[affiliation={2}]{Mateusz}{Kowalewski}
\author[affiliation={2}]{Leonid}{Pavlovskyi}
\author[affiliation={2}]{Sebastian}{Jurczak}
\author[affiliation={2}]{Anna-Mariia}{Vitkovska}
\author[affiliation={3}, orcid=0000-0001-6227-0960]{Zuzanna}{Miodonska}
\author[affiliation={4}, orcid=0000-0002-9754-6282]{Natalia}{Mocko}
\author[affiliation={3}, orcid=0000-0002-1770-6152]{Michal}{Krecichwost}

\address{
    $^1$ AGH University of Krakow, Cracow, Poland \\
    $^2$ SoftServe, Cracow, Poland \\
    $^3$ Department of Biomedical Engineering, Silesian University of Technology, Poland                          \\
    $^4$ Institute of Linguistics, Faculty of Humanities, University of Silesia in Katowice, Poland  
}

\email{miloszdudek@agh.edu.pl, hemmer@agh.edu.pl, kwarciak@agh.edu.pl, mstro@softserveinc.com, mpens@softserveinc.com, mkowale@softserveinc.com, lpavlov@softserveinc.com, sjurc@softserveinc.com, avitk@softserveinc.com, zuzanna.miodonska@polsl.pl, natalia.mocko@us.edu.pl, michal.krecichwost@polsl.pl}

\begin{document}
\maketitle

\begin{abstract}
Early identification of speech sound errors in children is often limited by access to specialists, motivating lightweight screening tools that can operate outside the clinic. We present a screening pipeline for Polish-speaking children focused on sibilant substitutions, coupling a wav2vec2-based CTC token recognizer with alignment-based error typing and a template-grounded caregiver assistant (screening, not diagnosis). On a held-out test set of 10 unseen children (559 utterances), the recognizer achieves 88.7\% exact sequence match. As a conservative screening proxy, we flag a mismatch when the system emits substitution-evidence (bracketed) tokens at the target segment, yielding 72.9\% precision, 61.4\% recall (F1=0.67) and a 2.7\% false-alarm rate (FPR on target-correct items). We describe the assistant's safety boundaries and outline a clinician-in-the-loop validation plan for future deployment.
\end{abstract}

\section{Introduction}
\label{sec:intro}

Speech sound disorders (SSDs) are among the most common developmental communication difficulties in early childhood and, when persistent, can affect intelligibility, participation, and later literacy-related outcomes \cite{waring2013classification}. In many settings, timely access to specialist assessment and therapy is constrained by limited capacity and long waiting times, motivating complementary approaches that support \emph{early screening} and caregiver-guided practice outside the clinic \cite{mcgill2021waitinglists,sugden2019homepractice}.

Polish is a particularly demanding language for automatic screening because it combines a dense consonantal inventory with frequent clusters. In Polish clinical practice, sibilants are often described as three series: \emph{sycz\k{a}cy} (e.g., \ipa{/s z \t{ts} \t{dz}/}), \emph{cisz\k{a}cy} (e.g., \ipa{/\textctc \textctz \t{t\textctc} \t{d\textctz}/}), and \emph{szumi\k{a}cy} (e.g., \ipa{/\:s \:z \t{t\:s} \t{d\:z}/}) \cite{zygis2010polish_fricatives,sage2024polish_sibilants}. Many referrals in Polish speech therapist practice concern errors in this space (often discussed under \emph{sigmatism} or \emph{lisping}), where substitutions (e.g., \ipa{/\:s/}$\rightarrow$\ipa{[s]}) and distortions may require different clinical handling \cite{sage2024polish_sibilants,pavsig2025}.

A practical home screening interaction must be short, repeatable, and focused on high-yield contrasts. We therefore build the screening loop around prompted words and syllables that probe place and manner contrasts relevant to Polish sibilants, including \emph{szafa}, \emph{szufelka}, \emph{sznurek}, \emph{stra\.zak}, \emph{czapka}, \emph{dziadek}, \emph{dzwonek}, \emph{zegar}, and minimal syllables such as \emph{ca/cia/cza}. The goal is not a fully general transcription system, but a sensitive and interpretable mechanism to flag likely errors and localize them at the token level.

Mispronunciation detection and diagnosis (MDD) has progressed from classical Goodness-of-Pronunciation pipelines toward neural architectures that aim to detect and localize errors at phoneme or feature levels \cite{shahin2025phonological_mdd,lounis2024mdd_systematic_review}. However, directly applying off-the-shelf ASR to child screening is problematic \cite{dudy2018_child_pronunciation_assessment,taekyung2024_asr_ssd_korean_children}. Child speech differs acoustically from adult speech (shorter vocal tracts and higher formants), shows higher within-speaker variability, and includes systematic non-canonical realizations that are part of development \cite{dudy2018_child_pronunciation_assessment,patel2024child_asr_improving}. Moreover, strong language-model priors in end-to-end ASR can normalize unusual realizations toward likely words, which is helpful for transcription but harmful for minimal-contrast error detection \cite{radford2022whisper}.

\begin{figure*}[t]
  \centering
  \includegraphics[width=0.8\textwidth]{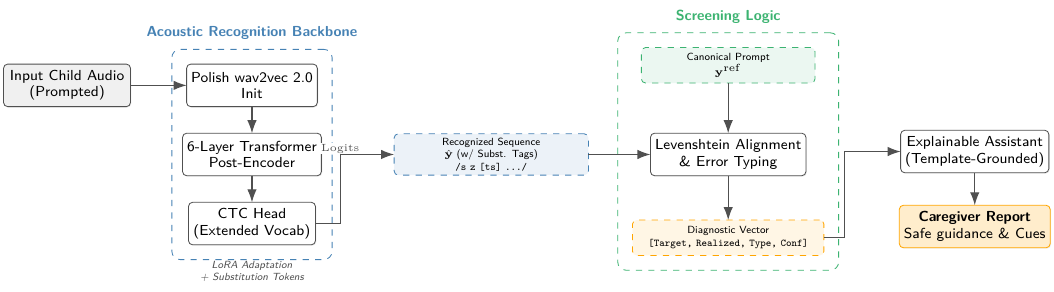}
  \caption{Overview of the screening loop. A wav2vec2-based recognizer outputs a production token sequence, which is aligned to the prompted canonical sequence to derive an interpretable screening vector. A caregiver-facing assistant maps this vector to template-grounded feedback with conservative escalation rules.}
  \label{fig:pipeline}
\end{figure*}

We therefore propose a caregiver-oriented screening loop that prioritizes phonetic sensitivity and \emph{operationally defined explainability} over general transcription quality (Figure~\ref{fig:pipeline}). In this paper, \emph{explainable} means: (i) the system produces an auditable token-level alignment and named error type, (ii) caregiver feedback is generated from a fixed, reviewable template inventory, and (iii) the system has explicit suppression/refusal rules under uncertainty. The acoustic module is a self-supervised speech encoder fine-tuned for Polish token recognition \cite{baevski2020wav2vec2,hsu2021hubert,peng2023e2e_mdd_transfer}. To make common sibilant substitutions explicit, we extend the token inventory with a small set of \emph{bracketed} IPA tokens that represent expert-judged substitution outcomes (e.g., ``closest to [\ipa{s}]''). On top of recognition, we perform alignment-based screening: recognized production tokens are aligned to the prompted canonical sequence to extract an interpretable diagnostic vector, which is then translated into caregiver-friendly feedback by an assistant constrained by templates \cite{shankar2025asr_llm_children}.

Importantly, we do not claim clinical diagnosis. We evaluate recognition and a conservative screening proxy on held-out children, and we present the assistant as a deployable design constrained by clinician-authored templates; full clinical and caregiver validation is part of our planned next step.

Our contributions are:
\begin{itemize}
  \item a substitution-aware token recognizer for Polish child speech built on wav2vec2 with a compact 6-layer Transformer post-encoder,
  \item a screening-oriented evaluation on held-out children that reports both recognition quality and conservative detection of target sibilant mismatches,
  \item a template-grounded assistant design that translates token-level screenings into caregiver-friendly feedback with explicit safety boundaries and a clinician-in-the-loop validation plan.
\end{itemize}

\section{Data}
\label{sec:data}
We use a proprietary corpus of prompted Polish child speech licensed for research \cite{pavsig2025}. Each utterance is associated with (i) the orthographic prompt, (ii) a canonical token sequence, and (iii) a token-level transcription of the child's production.

Aggregate, privacy-preserving dataset description:
\begin{itemize}
  \item Participants: 201 children (107F/94M), age 4--8 years (mean 76.2 months, SD 8.6), native Polish speakers recruited from schools/kindergartens in Poland.
  \item Recording conditions: On-site SLT examinations using a dedicated multimodal device: 15-channel audio (44.1~kHz/16-bit) and dual-camera oral-region video (30~fps). Quiet educational environments.
  \item Prompt inventory: 51 words and 12 logotomes covering all 12 Polish sibilants, elicited via picture naming and repetition. Total: 12{,}830 word/logotome segments and 53{,}951 phoneme segments.
  \item Annotation protocol: Manual segmentation into words and phonemes with IPA labels. For common sibilant substitutions, speech and language therapy experts provided \emph{bracketed} substitution evidence (Section~\ref{sec:token_repr}).

 % \item Ethics/consent and privacy: Bioethics approval obtained; parental consent and child assent collected: Bioethics Committee for Scientific Research at the Jerzy Kukuczka University of Physical Education in Katowice, Poland (Decision No. 3/2021)
\end{itemize}

\section{Method}
\label{sec:method}

\subsection{Token representation and substitution markers}
\label{sec:token_repr}
We represent pronunciations as sequences of \emph{space-delimited IPA-like tokens} for CTC training and decoding. Tokens are primarily phoneme-like segmental units; importantly, Polish affricates are represented as single tied symbols (e.g., \ipa{\t{ts}}, \ipa{\t{dz}}) in line with Polish phonology, even though they are written with digraphs in orthography (e.g., \emph{c, dz, cz, dż}). Tokenization choices are driven by \emph{CTC stability} on very short prompts: in a small subset of items we keep a fused consonant--vowel label used in the corpus annotations to reduce boundary jitter, hence we refer to \emph{tokens} rather than strictly \emph{phonemes} and report token-sequence length statistics in Section~\ref{sec:exp}.

To retain evidence of clinically frequent sibilant substitutions, we augment the label set with a small number of \emph{bracketed} IPA tokens (e.g., \ipatok{s}, \ipatok{\:s}, \ipatok{\textctc}, \ipatok{\t{ts}}, \ipatok{\t{t\:s}}). These markers encode an expert-assigned \enquote{closest-match} label for a substituted target and are treated as separate classes during training and inference.

Bracketed tokens are used \emph{conservatively} at screening time: we treat a bracketed token at the target position as explicit evidence of a likely deviation, while non-bracketed but non-canonical outputs are handled as uncertain (e.g., request a repeat recording or defer to speech-language expert review). This choice reduces false alarms in caregiver-facing use, and aligns with the assistant safety rules. The scheme covers 12 common outcomes, grouped by the three Polish sibilant series (orthographic forms in italics): (\emph{s, z, c, dz}) $\rightarrow$ \ipatok{s}, \ipatok{z}, \ipatok{\t{ts}}, \ipatok{\t{dz}}; (\emph{\'s, \'z, \'c, \'d\'z}) $\rightarrow$ \ipatok{\textctc}, \ipatok{\textctz}, \ipatok{\t{t\textctc}}, \ipatok{\t{d\textctz}}; (\emph{sz, \.z/rz, cz, d\.z}) $\rightarrow$ \ipatok{\:s}, \ipatok{\:z}, \ipatok{\t{t\:s}}, \ipatok{\t{d\:z}}.

\subsection{Acoustic model}
We formulate token recognition as CTC-based sequence labeling \cite{graves2006ctc}. The model consists of:
(i) a wav2vec2 encoder initialized from a Polish checkpoint: \emph{jonatasgrosman/wav2vec2-large-xlsr-53-polish},
(ii) a 6-layer Transformer post-encoder that refines temporal context, and
(iii) a linear projection to the token vocabulary trained with the CTC objective.
We add the post-encoder to stabilize CTC emissions in sibilant-heavy segments and to provide task-specific contextualization. In the controlled ablation reported in Table~\ref{tab:baselines}, removing the post-encoder reduces test exact sequence match from 88.7\% to 84.5\% (-4.2 pp) under otherwise identical training and decoding settings.

\subsection{Parameter-efficient adaptation}
To adapt the model to child speech and the extended token inventory, we combine LoRA-based adaptation \cite{hu2022lora} with partial unfreezing. Specifically, LoRA modules are injected into attention projections and feed-forward sublayers with rank $r{=}32$, $\alpha{=}64$, and dropout 0.1, and we additionally unfreeze the last 6 encoder layers. This yields 33.3\% trainable parameters (119.7M out of 359.6M). We describe this as \emph{efficient relative to full fine-tuning} rather than fully lightweight.

\subsection{Alignment-based screening and confidence}
Given a prompted canonical sequence $\mathbf{y}^{\mathrm{ref}}$ and a recognized production sequence $\hat{\mathbf{y}}$, we compute a minimum-edit (Levenshtein) alignment over tokens. We interpret alignment operations as:
\begin{itemize}
  \item Match: token-level agreement.
  \item Substitution: candidate mispronunciation event (or bracket evidence).
  \item Insertion/Deletion: candidate epenthesis/omission (reported conservatively).
\end{itemize}
For substitutions involving the sibilant/affricate focus set (Section~\ref{sec:token_repr}), we map the target--realized pair to an interpretable category such as (i) a place-of-articulation shift within the sibilant space, (ii) a voicing alternation, or (iii) a manner-of-articulation mismatch (affrication vs.\ frication). Substitutions outside this focus set (e.g., non-sibilant tokens such as \ipa{/f/} or \ipa{/v/}) are reported as generic mismatches without fine-grained typing, keeping the screening logic conservative. This yields a compact diagnostic vector:
\texttt{prompt}, \texttt{target}, \texttt{realized}, \texttt{type}, \texttt{position}, \texttt{confidence}.

To calculate confidence let $p_t(k)$ denote the frame-level posterior probability for token $k$ at frame $t$ under the CTC model. For an aligned target position $i$ with realized token $\hat{y}_i$, we define a confidence score as the mean posterior of $\hat{y}_i$ over the set of frames assigned to it by greedy CTC collapse:
\[
c_i = \frac{1}{\lvert T_i\rvert}\sum_{t\in T_i} p_t(\hat{y}_i), \qquad c_i \in [0,1].
\]
Here, $T_i$ denotes the set of non-blank CTC frames assigned to the aligned realized token $\hat{y}_i$ during greedy decoding, after repeated-label collapse and blank removal.
The assistant uses $c_i$ only as a \emph{suppression} signal (e.g., request a repeat recording when $c_i<\tau$); we do not claim calibrated probabilities.

For each prompt we define a small, manually curated mapping $p \mapsto F(p)$ that specifies the index/indices of the primary sibilant/affricate target(s) within the canonical token string.
For prompts with multiple focus targets, we mark the utterance as a mismatch if \emph{any} focus target is bracket-labeled in the reference (and analogously for predictions).

In our prompt design, most items isolate one primary sibilant target, making alignments effectively unambiguous. We therefore use uniform edit costs and perform linguistic error typing \emph{after} alignment via a deterministic target--realized mapping. %Phonologically weighted edit costs are a natural extension for broader prompt sets and multi-segment targets (future work).

\section{Experimental setup}
\label{sec:exp}
We train on 10{,}508 utterances and validate on 1{,}170 utterances, with a speaker-disjoint held-out test set containing 559 utterances from 10 unseen children. All splits share the same \emph{fixed prompt inventory} (words and syllables) by design, reflecting the intended screening setting where a child is asked to repeat a predefined set of diagnostic prompts. Consequently, our evaluation primarily measures \emph{generalization to unseen speakers} rather than open-vocabulary generalization to unseen lexical items.

Audio is converted to mono (channel averaging if needed) and resampled to 16~kHz to match the input interface of the SSL encoders used in this work. Recordings are single-word/syllable productions (typically around 1~s).

Because prompts are short (words/syllables), token sequences are also short by design: on the test set, reference strings have mean token length 3.66 (median 4, min 1, max 7). Reporting this distribution helps interpret exact-match metrics for short screening prompts.

We train with mixed precision and early stopping (patience 50; max 200 epochs). Training stops for wav2vec2 model with post-encoder layers and bracket tokens, with the best checkpoint at epoch 54 selected by validation performance (val WER 5.51\%, val CER 3.29\%). To contextualize the encoder choice, we additionally train a WavLM-Base baseline by \emph{replacing only the SSL encoder} while keeping the token vocabulary, post-encoder, LoRA configuration, training schedule, and greedy decoding unchanged (Table~\ref{tab:baselines}). This baseline is reported for comparison and is not used in the proposed screening pipeline. For decoding, we use greedy CTC decoding (argmax over logits) without an external language model in order to avoid lexical priors that could mask minimal-contrast errors \cite{radford2022whisper}.

\section{Results}
\label{sec:results}

\begin{table}[t]
\caption{Main recognition results on held-out test children (559 utterances).}
\label{tab:main_results}
\centering
\begin{tabular}{lr}
\toprule
Metric & Value \\
\midrule
Exact sequence match & 88.7\% (496/559) \\
Token accuracy & 95.0\% \\
WER (token string) & 5.95\% \\
CER (token string) & 4.09\% \\
\bottomrule
\end{tabular}
\end{table}

\begin{table}[t]
\caption{Baselines and ablations (held-out test set).}
\label{tab:baselines}
\centering
\small
\setlength{\tabcolsep}{6pt}
\renewcommand{\arraystretch}{1.05}

\begin{tabularx}{\columnwidth}{@{}Xccc@{}}
\toprule
Model & Exact (\%) & Token Acc. (\%) & Screening F1 \\
\midrule
\makecell[l]{wav2vec2 + \\ post-enc + \\ bracket tokens} & 88.7 & 95.0 & 0.67 \\
\makecell[l]{wav2vec2 + \\ bracket tokens} & 84.5 & 90.2 & 0.62 \\
\makecell[l]{WavLM + \\ post-enc + \\ bracket tokens} & 78.6 & 86.6 & 0.54 \\
\bottomrule
\end{tabularx}
\end{table}

Table~\ref{tab:main_results} summarizes headline recognition performance. Table~\ref{tab:baselines} provides a compact set of baselines/ablations to contextualize the design choices (post-encoder and bracket supervision). We emphasize exact sequence match because a 1:1 match enables stable downstream alignment and error typing without additional heuristics. Validation metrics were comparable to test (val WER 5.51\% vs.\ test WER 5.95\%), suggesting limited overfitting under the current prompt set.

Across the 10 test children (prompted with a fixed inventory of 64 prompts/items; some recordings may be missing), exact match ranges from 72.6\% to 100\% (mean 89.5\% $\pm$ 9.4\%). Utterance-level bootstrap confidence intervals are: exact match 95\% CI [86.1, 91.2] and token accuracy 95\% CI [93.8, 96.3]. Because utterances are clustered by child, we additionally report a \emph{cluster bootstrap by child} (resampling children with replacement and keeping all their utterances). With only 10 test children, these CIs are wider and better reflect speaker-generalization uncertainty: Exact match 95\% CI [83.8, 93.2], token accuracy 95\% CI [92.7, 97.0]. This quantifies uncertainty given the small held-out cohort. 

\subsection{Screening proxy evaluation (target sibilant mismatch detection)}
Recognition quality alone does not guarantee screening utility. We therefore evaluate a minimal screening decision focused on the primary sibilant target in each prompt. In our corpus, deviations at the target segment are annotated using bracketed evidence tokens, so we are using the rule that if there is a flag mismatch we have recognized token at target position that is bracketed. On the held-out test set (the resulting confusion matrix can be seen in Table ~\ref{tab:screening_cm} where the child has pronounced particular word), this rule yields 72.9\% precision and 61.4\% recall (F1=0.67) for detecting target mismatches, with a low false-alarm rate of 2.7\% on target-correct items. When a mismatch is flagged (true positives), the predicted bracket class matches the reference bracket label in 85.7\% of cases, supporting interpretable error typing.

\begin{table}[t]
\caption{Screening proxy confusion matrix on the held-out test set (focus targets only).}
\label{tab:screening_cm}
\centering
\small
\setlength{\tabcolsep}{6pt}
\renewcommand{\arraystretch}{1.05}
\begin{tabular}{lcc}
\toprule
 & $\hat{y}=1$ (flag) & $\hat{y}=0$ (no flag) \\
\midrule
$y=1$ (ref mismatch) & TP = \textbf{35} & FN = \textbf{22} \\
$y=0$ (ref correct)  & FP = \textbf{13} & TN = \textbf{489} \\
\bottomrule
\end{tabular}
\end{table}

\subsection{Residual error analysis on sibilant units}
To focus on clinically relevant contrasts, we analyze errors at a \emph{sibilant unit level} by extracting the 12 symbols (canonical and bracketed) from each token string and aligning these unit sequences with a minimum-edit algorithm. Among the 63 non-exact utterances, 58 (92.1\%) contain exactly one unit-level substitution, indicating that most recognition errors are localized to a single sibilant decision. Of these, 35/58 (60.3\%) are canonical-bracket confusions (one side bracketed), consistent with the role of bracket tokens as graded substitution evidence.

% \begin{table}[t]
% \caption{Residual unit-level error breakdown on the test set (63 non-exact utterances).}
% \label{tab:error_breakdown}
% \centering
% \small
% \setlength{\tabcolsep}{4pt}
% \renewcommand{\arraystretch}{1.05}
% \begin{tabularx}{\columnwidth}{@{}X r@{}}
% \toprule
% Statistic & Value \\
% \midrule
% Non-exact utterances & 63/559 (11.3\%) \\
% Utterances with exactly one sibilant/affricate substitution & 58/63 (92.1\%) \\
% Of which canonical$\leftrightarrow$bracket confusions & 35/58 (60.3\%) \\
% \midrule
% Top confusions (ref$\rightarrow$pred) & count \\
% \hspace{1em}\ipa{\:s} $\rightarrow$ \ipatok{s} & 4 \\
% \hspace{1em}\ipa{\:s} $\rightarrow$ \ipatok{\t{t\:s}} & 3 \\
% \hspace{1em}\ipatok{s} $\rightarrow$ \ipa{\:z} & 3 \\
% \bottomrule
% \end{tabularx}
% \end{table}

%Figure~\ref{fig:confusions} summarizes the most frequent patterns. False alarms are dominated by \ipa{\:s}$\rightarrow$\ipatok{s} and \ipa{\:s}$\rightarrow$\ipatok{\t{t\:s}}, while misses on substitution-labeled segments are dominated by \ipatok{s}$\rightarrow$\ipa{\:z} and \ipatok{\t{ts}}$\rightarrow$\ipa{z}.

% JAK COŚ TO MOŻEMY WYRZUCIĆ TE BŁĘDY
% \begin{figure*}[t]
 %  \centering
%   \includegraphics[width=0.7\textwidth]{figures/confusions.pdf}
%   \caption{Most frequent sibilant unit-level confusions on non-exact utterances. Left: false substitution evidence on canonical reference units. Right: errors on substitution-evidence (bracketed) reference units.}
 %  \label{fig:confusions}
% \end{figure*}

\section{Explainable assistant design}
\label{sec:assistant}

The recognition and alignment modules provide token-level screening evidence, but caregiver value depends on presenting it in a safe and interpretable form. The assistant consumes the screening vector \texttt{[target, realized, type, position, confidence]} and generates a short caregiver report grounded in clinician-authored templates.

\noindent\textbf{Caregiver-facing view.}
For each prompt, the caregiver sees the word in orthography, a match/likely-mismatch flag (with a \emph{repeat} suggestion under low confidence), a plain-language localization (e.g., \enquote{at the start}), the minimal contrast in orthography (e.g., \emph{s} vs.\ \emph{sz}), and an optional brief practice cue. IPA strings are hidden by default; clinicians can audit the underlying alignment, token outputs, error type, and confidence.

\noindent\textbf{Template grounding and safety.}
Messages are produced by filling a fixed, reviewable template inventory keyed by \texttt{type}. Templates include conservative escalation rules and refusal constraints (no diagnosis, no medical claims, no guarantees). Under low confidence ($c_i<\tau$) or inconsistent evidence, the assistant requests re-recording instead of giving detailed advice \cite{shankar2025asr_llm_children}.

\begin{table}[t]
\caption{End-to-end example of the screening loop (illustrative test utterance).}
\label{tab:e2e_example}
\centering
\small
\setlength{\tabcolsep}{3pt}
\renewcommand{\arraystretch}{1.05}
\begin{tabularx}{\columnwidth}{@{}l X@{}}
\toprule
Prompt & \emph{sznurek} \\
Target (IPA) & \ipa{/\,\:s n u r E k\,/} \\
Recognized (IPA) & \ipa{[s] n u r E k} \\
Screening entry & \texttt{target=} \ipa{\:s}, \texttt{realized=} \ipatok{s}, \texttt{type=place shift}, \texttt{pos=1}, \texttt{conf=0.91} \\
Caregiver message & ``At the start of \emph{sznurek}, it sounded closer to \emph{s} than \emph{sz}. Try repeating slowly and aiming for a slightly more retracted tongue position than for \emph{s}. If this pattern persists over days/weeks, consult a speech-language pathologist.'' \\
\bottomrule
\end{tabularx}
\end{table}

\noindent\textbf{Report aggregation.}
Across prompts, the system can aggregate flags into a compact summary (top recurring contrasts + short practice plan), while keeping the interaction within a screening scope.

\section{Discussion and limitations}
\label{sec:disc}

The results suggest that SSL-based token recognition can provide a practical backbone for caregiver-oriented screening in Polish. By designing the pipeline around prompted contrasts and alignment to the canonical target, we avoid reliance on word-level language-model priors that may mask minimal-contrast errors \cite{radford2022whisper}. The high exact-match rate supports stable downstream alignment; nevertheless, screening performance depends on which deviations are detected versus missed, and on controlling false alarms.

Bracketed evidence labels reflect an expert interpretation of a non-canonical realization. Our error analysis shows that many residual errors involve canonical-bracket confusions. This motivates future exploration of alternative formulations (e.g., a two-head model that predicts a base phone and a separate deviation flag, or label smoothing between canonical and bracketed counterparts) to better reflect the graded nature of these contrasts.

Because the intended use case relies on a fixed diagnostic prompt list, we evaluate speaker-disjoint generalization on the same prompt inventory across splits; generalization to unseen lexical items is left for future work.

The held-out evaluation covers 10 unseen children and 559 utterances; broader evaluation across ages, severities and recording conditions is needed before deployment. We report per-child variability and bootstrap confidence intervals, and we plan to add age-stratified analyses.

While the assistant is template-grounded by design, we do not yet present user studies with caregivers or clinicians. Evaluating safety, clarity and risk of misinterpretation (screening vs.\ diagnosis) is essential future work.

\section{Conclusion}
\label{sec:conclusion}

We presented a phoneme-level mispronunciation screening loop for Polish-speaking children based on wav2vec2 token recognition, alignment-based screening, and a template-grounded caregiver-facing assistant. The recognizer achieves 88.7\% exact sequence match on unseen children, supporting stable token-to-token alignment. We additionally report a conservative screening proxy evaluation based on bracketed substitution evidence, yielding low false-alarm rates and interpretable error typing. Future work will expand coverage beyond substitutions, validate diagnostic utility against clinician outcomes, and evaluate assistant feedback quality and safety in caregiver-facing studies.

\section{Acknowledgments}
This work was supported by the National Centre for Research and Development (NCBR), Poland, under research project No. 0179/L-15/2024, entitled ``Speech therapy computer system for recording and analyzing multimodal articulation data using the 4D speaker model and deep learning (SpeechCAD).'' 
This study was approved by the Bioethics Committee for Scientific Research at the Jerzy Kukuczka University of Physical Education in Katowice, Poland (Decision No. 3/2021). Written parental consent and child assent were obtained for all participants.

\section{Generative AI Use Disclosure}
Generative AI tools were used to assist with language editing, polishing, and local wording. They were not used to produce a significant part of the manuscript. All authors reviewed, edited, and verified the final manuscript and remain responsible for its content.

\bibliographystyle{IEEEtran}
\bibliography{mybib}

\end{document}